# Perturbation-induced granular fluidization as a model for remote earthquake triggering

**Authors**
Kasra Farain[1] and Daniel Bonn,[1]*

**Affiliations**
[1]Van der Waals–Zeeman Institute, Institute of Physics, University of Amsterdam; Science Park 904, 1098 XH Amsterdam, Netherlands.
*d.bonn@uva.nl

**Abstract**
Studying the effect of mechanical perturbations on granular systems is crucial for understanding soil stability, avalanches, and earthquakes. We investigate a granular system as a laboratory proxy for fault gouge. When subjected to a slow shear, granular materials typically exhibit a stress overshoot before reaching a steady state. We find that short seismic pulses can reset a granular system flowing in steady state so that the stress overshoot is regenerated. This new feature is shown to determine the stability of the granular system under different applied stresses in the wake of a perturbation pulse and the resulting dynamics when it fails. Using an analytical aging-rejuvenation model for describing the overshoot response, we show that our laboratory-derived theoretical framework, can quantitatively explain data from two fault slip events triggered by seismic waves.

**Introduction**
Granular materials can be solid-like with the ability to resist imposed stresses up to a maximum value called the yield stress. Above this stress, the materials start flowing more or less like liquids, as seen in an hourglass (*1, 2*). The transition from solid to liquid-like state and the resulting rheological properties of granular materials are very sensitive to mechanical vibrations (*3, 4*). Dynamic stresses much smaller than the yield stress of the material can dramatically weaken it by reducing the stresses at the contacts between particles. Once weakened, the whole granular assembly can start moving under a smaller overall shear stress and experience a sudden fluidization. For instance, if the sand flowing through an hourglass jams, a small tap is enough to make it start flowing again. Such vibration induced weakening of granular media evidently has important implications for avalanche and earthquake triggering; notably it has been proposed as a mechanism by which seismic waves of some earthquakes can activate additional events at large distances where the static stress changes are negligible (*5-9*). A central question that remains unanswered however is whether the failure induced by vibrations can be described or predicted based on the rheological properties of unperturbed granular media. Additionally, a comprehensive theoretical framework of the fluidization process should account for the changes in the rheology of the granular system *after* the perturbation, leading to the eventual recovery of the system's strength.

In this study, we investigate a granular medium's responses under different loading conditions to a short-lived (about 100 ms) pulse perturbation. We show that when a granular material is under a constant shear rate, the perturbation pulse leads to an overshoot in the stress response of the material as the perturbed random granular configuration returns to a steady-state flow configuration again. The height of this stress overshoot varies depending on the magnitude of the perturbation pulse, suggesting that the (static) yield stress is not a constant material property and depends, in addition to existing environmental noise, on the perturbation history of the material. Furthermore, we find that a perturbation pulse must be of sufficient magnitude compared to the fault's static shear stress to activate a frictional slip on it. In this case, the resulting dynamics are then determined by the relationship between applied stress and steady-state resistance: if the stress exceeds the steady-state resistance, the flow will accelerate indefinitely; otherwise, it will come to a halt after a jump in the shear deformation. Our model experiments are analogous to situations where faults in an aseismic slip or under constant static stresses are impacted by seismic waves from far-field earthquakes.

**Stress overshoot**

We first study the yielding behavior of a non-agitated granular material by measuring its stress response in a quasi-stationary flow under a small imposed shear rate. We connect a rheometer to a thick cylindrical tube resting on a layer of hard, spherical polymethyl methacrylate (PMMA) particles (~ 40 μm diameter) (Fig. 1A). The rheometer applies a constant rotational speed to the tube, while measuring the material's friction $F$ (Fig. 1B). As the granular layer under the tube begins to flow, the measured friction exhibits a long continuous overshoot with a peak $F_{peak}$ that is significantly higher than the resistance $F_{SS}$ that it maintains in steady state ($t > 400$ s). Under an applied stress $\Sigma$, in the absence of any external perturbation, the granular material yields to flow if $\Sigma > F_{peak}$, and remains stable perpetually for any $\Sigma$ smaller than $F_{peak}$ (*10*).

The stress overshoot of the onset of granular flow arises when the system transforms, under a constant applied shear rate, from a rest configuration to a steady-state configuration. These packing configurations cannot be the same as the stresses in each configuration are different (Fig. 1B, insets). In the rest configuration, the resultant of the microscopic forces is a normal vector, equal and opposite to the external normal load on the granular material. However, in the flowing steady-state configuration, the granular material must also provide shear resistance $F_{SS}$.

**Perturbations on sheared granular systems**

A particular configuration of granular material can collapse when subjected to a mechanical perturbation pulse, resulting in a new random state. Essentially, this denotes that the aging and frictional sliding memory of the granular packing can be erased by mechanical perturbations. This is analogous to heating a polymer glass above the glass transition temperature and then cooling it down again to achieve a glassy state, which eliminates the thermal history of the polymer. In Fig. 2A, the first peak on the left shows the expected overshoot of granular friction during the transition from an initial rest configuration to steady-state flow under a constant applied shear rate. After the granular flow reaches a steady state, at $t = 305$ s (indicated by a black arrow in Fig. 2A), an elastic wave pulse (Fig. 2B) perturbs the granular flow, causing an instantaneous drop in frictional resistance at the sliding interface. The interface then exhibits friction dynamics similar to those observed during the rest to flow transition. The duration of the wave pulse is about 100 ms (Fig. 2B), much shorter than the time scale for the continuous transient dynamics; however, the weakening effect of the wave pulse persists in the granular system long after the pulse has dissipated (Fig. 2C). The observed structural weakness in the granular system is due to aging-rejuvenation dynamics of the granular packing that emerge in a solid-like quasi-static state. This differs from the reduction of granular friction under continuous acoustic waves previously suggested as an explanation for some weak faults, where the agitated particles are constantly colliding (*3, 4, 7-9*). In our experiments, passing wave pulses cause momentary fluidization of the jammed state, leading to the reorganization of particles in a random configuration and removal of previous frictional aging and sliding memory. The system then returns to quasi-static frictional contacts governed by aging-rejuvenation dynamics. The third and fourth peaks in Fig. 2A show that smaller-amplitude wave pulses result in smaller drops in friction due to partial fluidization and removal of flow history. It is also worth noting that the duration of the continuous frictional dynamics depends on the sliding speed, as the aging-rejuvenation evolution proceeds with the sliding distance (*10*).

**Perturbations on stressed granular systems**

In natural fault systems, the granular fault material is usually subjected to a constant static stress from the surrounding environment when a wave pulse impacts the fault. During the passage of the pulse, the granular material can fluidize, and the friction overshoot is eliminated. If the static stress $\Sigma$ is greater than the steady-state friction $F_{SS}$, a flow is initiated that can accelerate unrestrictedly (Fig. 3A). But if $\Sigma < F_{SS}$, the deformations will not be large as the granular medium regains its strength after some time (Fig. 3B). It is important to note that, in natural faults, the slip is also constrained by the amount of stored elastic energy on the fault that is available to be released. As slip occurs on a natural fault, the stored elastic energy that supports the motion decreases until it is insufficient to support further slip, eventually causing the fault to become locked again. However, not all seismic waves that impinge on a stressed fault are powerful enough to trigger an earthquake. The threshold intensity required to induce a fault gouge failure depends on the static stress acting on the fault. If the friction drop resulting from the wave pulse impact does not fall below the applied static shear stress, the granular system remains stable, as shown in Fig. 3C. Under these circumstances, the observed deformations are typically only a small fraction of a particle size (several tens of nanometers, compared to 40 micrometers), and do not represent a failure of the granular structure.

**Comparison to natural triggered events**

Remote earthquake triggering was first documented after the 1992 magnitude 7.3 Landers earthquake (*5*). This event triggered additional earthquakes at distances up to 1250 km from the main shock. As both calculated and observed static stress changes at distances over 300 km were negligible (falling below the daily tidal stress changes), it was concluded that the propagating seismic waves must have caused the remote triggering. We now compare our laboratory-derived fault triggering and evolution principles with the post-Landers activity detected at the Long Valley caldera, which is the only site triggered by the Landers mainshock for which a continuous record of deformation exists (*5*). Strain changes measured using a borehole dilatometer displayed a small and immediate compression of about $3\times10^{-9}$ between the Landers compressional (P) and shear (S) waves. Next, a slower-growing compressional strain developed, reaching about $2\times10^{-7}$ within the first 5 days after the Landers event, gradually decaying to background level in the following weeks. This corresponds to the post-perturbation non-monotonic friction evolution observed in our granular system (Fig. 4A). It should be noted that the measured dilatational strain $\sigma$ is directly related to the stress $\theta$ by $\sigma \approx k\theta$, where $k$ is the bulk modulus (*5*). The overshoot response can be understood by a simple first-order differential equation which reflects the competition between aging of the granular packing and a 'rejuvenation' effect caused by shearing (*10*):

$$\frac{dF}{dt} = \frac{a}{t} - \alpha v (F - F_{SS}),  \qquad (1)$$

where $a$ and $\alpha$ are material-dependent constants. Solutions are of the form:

$$F(t) = a e^{-\alpha v t} \int_{t_0}^{t} \frac{e^{\alpha v t'}}{t'} dt' + F_{SS}(1 - e^{-\alpha v t}), \qquad (2)$$

where the lower bound of the integral $t_0$ denotes the initial configurational state of the granular pack that the evolution presented by Eq. 2 initiates from. The latter is similar to the state variable in the empirical rate-and-state friction equations that have extensively been used to study rock friction dynamics (*11*). However, it is important to note that the observed non-monotonic evolution cannot be described by the usual rate-and-state friction model: in that model, all stress evolutions are monotonic.

A second real-world example of triggered non-monotonic evolution is provided by pressure data from multiple levels in an offshore borehole drilled into Nankai subduction zone (site 808), southwestern Japan (Fig. 4B) (*12*). In July 2003, following some low-frequency seismic events, a pressure rise was observed at different levels of the borehole simultaneously. The high level of concurrency of the onset of pressure pulses over a 500 m vertical distance, and also in comparison with signals observed at a laterally distanced site, indicates that hydraulic diffusion cannot be the mechanism by which the signal transmission occurred (*12*). Instead, these observations are consistent with the seismic wave triggering described above. Furthermore, the maximum pressure at each level was reached after different times ranging from 6 hours to more than one day. When pressure pulses start off together, what defines this unharmony of their maxima? From Eq. 1, the maximum friction force $F_m$, where $\frac{dF}{dt} = 0$, and the time between onset and maximum, $t_m$, are related as

$$\frac{1}{t_m} = \frac{\alpha v F_{SS}}{a} \left( \frac{F_m}{F_{SS}} - 1 \right). \qquad (3)$$

Our experiments indeed indicate that $\frac{\alpha v F_{SS}}{a}$ is a constant (*10*). Additionally, $F_{SS}$ is proportional to the normal load on the granular material according to Amontons' law (*13*). This normal load at each level of the Nankai borehole is specified by the amount of the granular solid material on that level or the depth of the level below the seafloor $D$. The depth of the seafloor itself is not important for our purpose because the fluid just imposes a buoyant force that essentially reduces the density of the granular material to $\rho_R - \rho_W$, where $\rho_R$ is the density of rock and $\rho_W$ is the sea water density. In the inset of Fig. 4B, we plot the maxima of the pressure pulses $P_m$ normalized by the depths $D$ at which the corresponding monitoring screens are positioned as a function of $\frac{1}{t_m}$. The result, excluding the largest pulse (measured at Nankai level 3, see Fig. 4B), confirms that the relation between the pressure maximum and its time of occurrence is determined by our aging-rejuvenation model, pointing also to similar material-dependent frictional properties at these levels.

**Discussion**

In summary, our laboratory-scale granular fault experiments and analytical description provide insights into remote earthquake triggering. We demonstrated that a seismic pulse can either activate a catastrophic failure on a fault that is under static stress, or initiate an aseismic non-monotonic deformation on a fault that is undergoing sliding motion controlled by its surrounding environment. Since earthquakes and tectonic phenomena follow scale-invariant laws (*14, 15*), our laboratory-scale frictional model can reproduce the main characteristics of aseismic slip and failure processes in a crustal fault. These findings differ from other dynamic triggering interpretations (*8, 9*), highlighting the crucial role of aging of the granular packing after a perturbation-induced change of the granular configuration. Specifically, the non-monotonicity described by Eq. 2 emerges due to this aging effect. Furthermore, our results suggest that granular material aging may offer an alternative explanation for quasi-static weak faults (*16-19*), as a fault gouge can remain weak for a long time after a seismic wave pulse due to the intrinsically slow aging process. The relevance of the slow evolution dynamics and different failure schemes, as presented here, to natural seismic phenomena has yet to be explored using continuous deformation data and aftershock occurrence statistics.

**Materials and Methods**

**Experimental setup**

Our experimental setup, as described in (*10*), comprises a thick cylindrical tube resting on a granular layer (~ 1 mm) of hard, spherical PMMA particles (~ 40 µm diameter) (Fig. 1A). The tube, made of aluminum, has an outer diameter of 29.7 mm, a wall thickness of 2.3 mm, and a height of 58 mm. To prevent wall slip, we sandblast both the tube cross-section and the bottom surface (also aluminum), resulting in a surface roughness of 2 – 3 µm (measured optically with a Keyence profilometer). We use a custom-made tool, connected to a rheometer (Anton Paar MCR 302), to rotate the tube around its symmetry axis. The rheometer is capable of moving the macroscopic tube with precision and uniformity down to 1 nm/s. This is crucial for capturing continuous instrumental dynamics, given the highly nonlinear nature of friction. It is also worth noting that relative motion between tectonic plates occurs at very low speeds, as low as a few centimeters per year. Therefore, our experimental setup's ability to simulate such slow, natural movements is particularly important for investigating the fundamental mechanics of fault motion. To ensure that the granular material experiences a constant load in the normal direction, we utilize low-friction bearings between the rheometer tool and the tube. This design feature mimics the natural fault scenario, where the granular fault zone material undergoes constant shearing under the weight of the overhead rock at depths of 10–20 km (*9*). Additionally, the use of low-friction bearings allows the tube to settle quickly on the granular material after a perturbation, avoiding damped oscillations that may occur in constrained granular materials where the normal load is applied through a spring and feedback system. These oscillations can obscure the short time scale response of the granular system.

**Perturbations**

In order to induce seismic waves on our laboratory-sized fault system, we generate short elastic wave pulses by bouncing a rubber ball off the bottom surface at a location 10 cm from the edge of the tube. These pulses are then captured with a piezo element (SparkFun Electronics, SEN-10293), which is affixed to the bottom surface on the opposite side of the tube, 3 cm away from the edge. The piezo element is connected to a digital oscilloscope (Tektronix, DPO3014) to record the data.

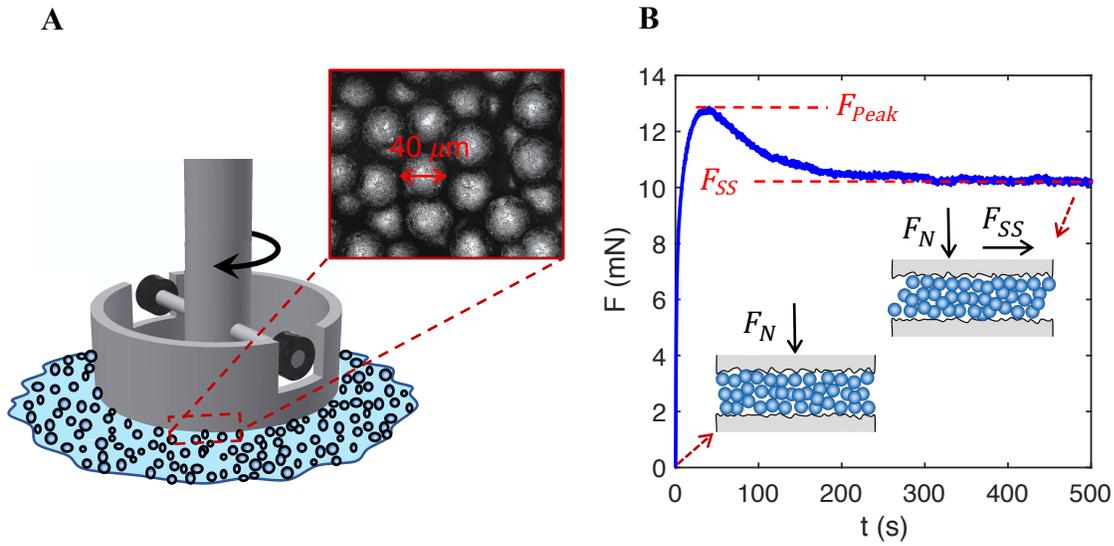

**Fig. 1. Experimental set-up and the stress overshoot of the onset of granular flows.**
**(A)** We use a rheometer to rotate a cylindrical tube that is resting on a granular layer around its symmetry axis. The rheometer can apply either a rotational torque or a rotational speed to the material, while measuring the other as the granular media's response. The inset shows an image of the microspheres taken by a Keyence optical profilometer. **(B)** The granular material starts from a random rest configuration (bottom left inset). Applying a constant shear rate causes it to start flowing. After a broad continuous stress overshoot, it arrives at a steady state with an aligned configuration (top right inset).

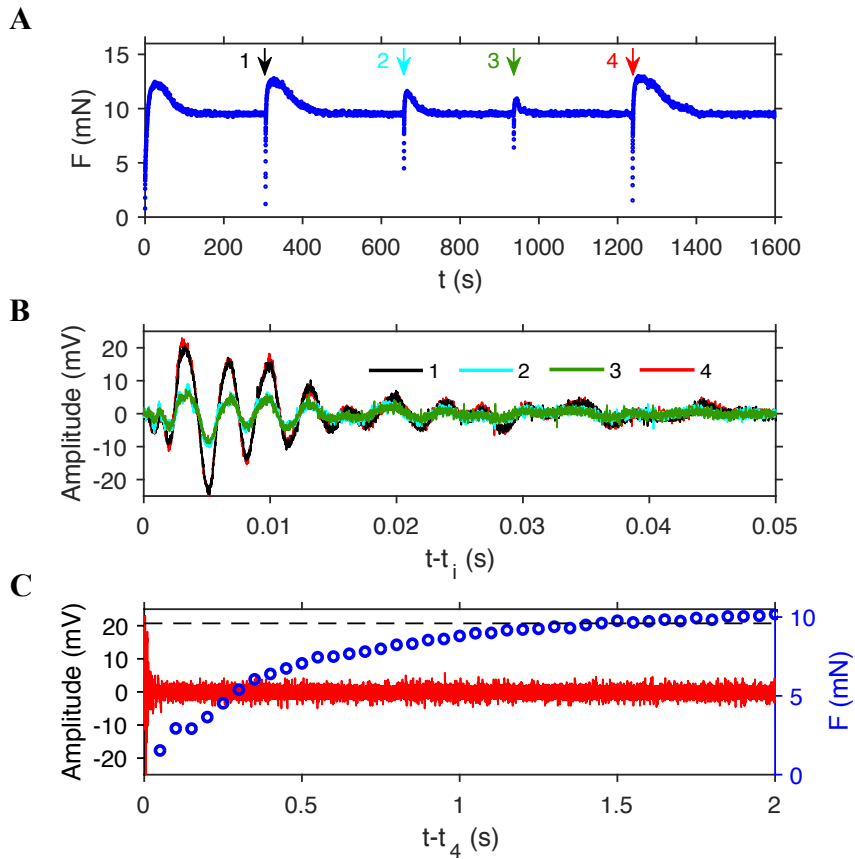

**Fig. 2. Response of laboratory-scale granular fault under a constant shear rate to a pulse perturbation.** (**A**), Frictional resistance of the granular interface as a function of time. Pulse perturbations of two different amplitudes are applied at times indicated by arrows 1 – 4. After each pulse perturbation, the granular friction quickly drops followed by a slow transient overshoot before the system reaches equilibrium. (**B**), Large- and smaller-amplitude pulse perturbations (shifted to start at $t = 0$) as recorded by a piezo transducer. The colors/numbers correspond to those of the arrows in **A**. (**C**), Enlarged view of the fourth applied perturbation (left axis) specified by the red arrow in **A** together with the subsequent friction drop and evolution (right axis). The black dashed line indicates the steady-state friction. Until about ~ 1.3 s after the perturbation pulse, the frictional strength is smaller than its steady-state value. This time increases with decreasing sliding speed (*10*).

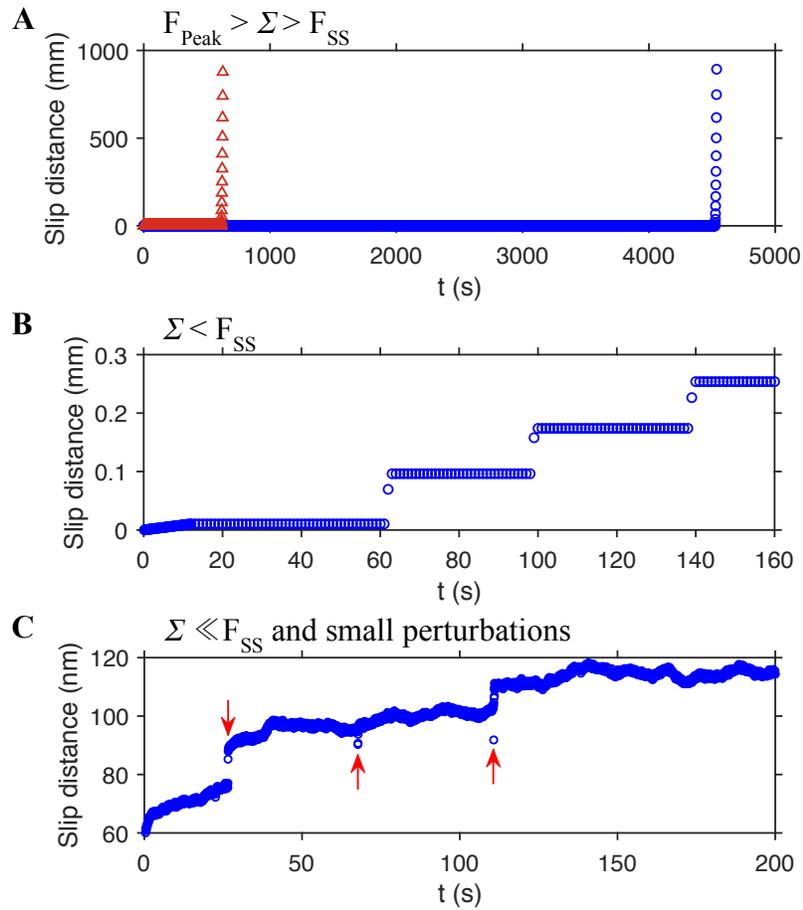

**Fig. 3. Typical responses of the granular fault under different shear forces $\Sigma$ to pulse perturbations of different intensity.** (**A**), At an applied shear force of 11.3 mN, which is above $F_{SS}$ = 9.5 mN but below the peak friction $F_{Peak}$, a perturbation takes place at time $t$ = 600 s. This triggers an accelerating sliding motion (red). In a second experiment (blue), the perturbation takes place at $t$ = 4500 s to show that the unperturbed system is stable for a long time. (**B**), At an applied shear stress of 9.1 mN, which is slightly below $F_{SS}$, the friction drop after the passage of perturbation triggers a jump of about 100 µm in the sliding direction. The perturbations in **A** and **B** are similar to black and red cases in Fig. 1. (**C**), At an applied shear stress of 1.5 mN, at three times indicated by arrows, small perturbations similar to the green and cyan cases in Fig. 1 are applied. Since these perturbations are not powerful enough to reduce the frictional strength to below the applied static stress, no failure occurs.

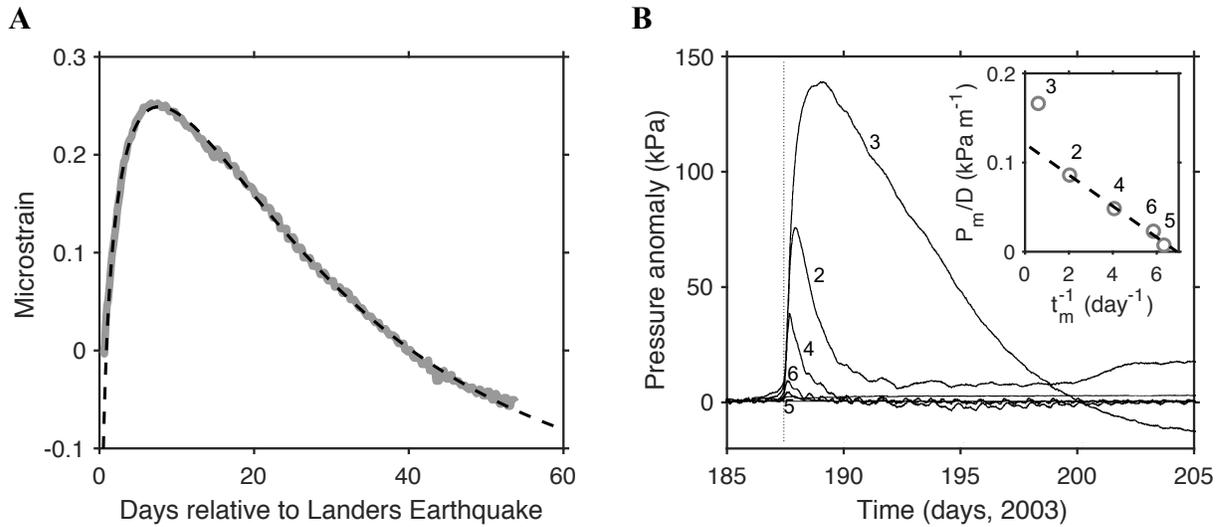

**Fig. 4. Comparison of the aging-rejuvenation model with borehole observations of strain and fluid pressure after earthquakes.** (**A**), Strain changes in Long Valley caldera after the 1992 magnitude 7.3 Landers earthquake (adapted from Fig. 6 in Ref. (*5*)). The dashed line is the best fit of Eq. 2. (**B**), Pressure data from different monitoring levels at site 808 in the Nankai subduction zone following a swarm of small very-low-frequency earthquakes (adapted from Fig. 5 in Ref. (*12*)). Pressure peaks 2, 3, 4, 5, and 6 were observed at depths $D$ = 878, 833, 787, 533, and 371 meters below seafloor, respectively. The inset shows the maximum pressure normalized by the screen depth $\frac{P_m}{D}$ as a function of the inverse of the time between start of pulses (indicated by a dotted line) and when the maximum pressure was reached $\frac{1}{t_m}$. The dashed line is a linear fit ignoring the first data point.